# Electro-optic modulation of coherent and incoherent mid-IR radiation in two-dimensional arrays

Jared Sisler, [1,†] Phillippe Pearson, [1,†] Michael Kelzenberg, [1] Andrei Faraon [1], Harry A. Atwater[1]*

[1] Thomas J. Watson Laboratories of Applied Physics, California Institute of Technology, Pasadena, California 91125, USA
† These authors contributed equally to this work
* Corresponding author: Harry A. Atwater, haa@caltech.edu

**Abstract**
Light in the mid-infrared (mid-IR) spans wavelengths from 3–8 $\mu$m and is important to many applications such as gas sensing and thermal imaging. Due to materials challenges, there is currently a lack of mid-IR reconfigurable optical elements. Here, we present an electrically addressable metasurface for modulation of coherent and incoherent mid-IR radiation in two spatial dimensions. Our device achieves optical modulation due to the field-effect free-carrier depletion in a lightly doped ($10^{19}$ cm$^{-3}$) film of indium-tin-oxide (ITO) coupled to a gap plasmon resonator. By addressing 32 individual elements across the metasurface, we first demonstrate tunable diffraction of coherently reflected mid-IR light. Next, we introduce a scalable perimeter-addressed driving scheme for tunable diffraction in two dimensions. Finally, we demonstrate modulated emissivity with spatially reconfigurable two-dimensional patterns at elevated temperatures. This work advances the development of solid-state reflective beam-steering devices in the mid-IR and manipulation of thermally emitted incoherent radiation.

**Introduction**

The mid-infrared (mid-IR) wavelength spectrum spans approximately 3–8 $\mu$m. It is of scientific and technological interest because it contains the peak emission for most thermal bodies as well as characteristic absorption lines of many molecules. It is critical for gas detection (*1*), thermal energy harvesting (*2*), medical imaging (*3*), and military sensing (*4*). Despite widespread interest, there is currently has a lack of mid-IR optical components (i.e. sources, detectors, modulators) compared to the near-IR and visible. This gap in technology is primarily caused by limitations of available materials. For example, due to the lower energy of mid-IR photons, detectors based on photoabsorption are more difficult to design. Additionally, liquid crystals commonly used in spatial light modulators for visible and near-IR wavelengths are highly absorptive in the mid-IR, requiring molecular redesign for compatibility (*5*). Because of this, current techniques for active wavefront manipulation of coherent mid-IR radiation usually rely upon mechanically tunable mirrors. As an example, conventional digital micromirror devices (DMDs) consist of a 2D array of micro-electromechanical system (MEMS)-tunable mirrors that can impart binary amplitude profiles onto reflected mid-IR light for projection (*6*), spectroscopy (*7*), and adaptive beam-forming (*8*). While mirror-based wavefront shaping devices have high efficiency and amplitude contrast, their switching speed is limited to tens of kHz (*9*), phase modulation is difficult to achieve (*10*), and their pixel pitch is typically limited to a few to tens of microns. Optical phased arrays (OPAs) have been demonstrated in the mid-IR as a possible method to achieve dynamic beam-steering (*11*), but their overall efficiency has been limited to a few percent and they are still restricted to modulation speeds in the 10s of kHz due to the use of the thermo-optic effect for active phase shifting elements. Thus, there is currently a need for devices capable of manipulating the wavefront of mid-IR light at high modulation rates (> MHz) with high efficiency and subwavelength spatial resolution.

Looking beyond the manipulation of coherent light, the mid-IR is also a useful wavelength range to study and engineer the properties of incoherently emitted thermal radiation. It has been shown that the radiation spectrum of a thermal body can be tailored by nanostructuring the material's surface to introduce optical resonances (*12–14*). Through similar techniques, the radiation pattern can be shaped to realize symmetric and asymmetric far-field diffraction (*15–17*). Additionally, various methods have been used to demonstrate dynamic modulation of spectral emissivity, such as the electrical modulation of the carrier density of a material when coupled to a mid-IR resonator (*18–21*). An interesting next step would be to demonstrate the tunable emissivity of a thermal emitter both in space and time. While there have been demonstrations of 2D arrays of electrically tunable thermal emitters, their modulation is based upon varying the local temperature (not the intrinsic emissivity) of each array element (*22, 23*). It has also been shown that the directivity of thermal emission can be electrically switched but, in this demonstration, the spatial emissivity profile across the device aperture was uniform (*24*). Additionally, a previous work used photogeneration of free carriers to locally modulate emissivity (*25*), but optical pumping is not always suitable to applications which need compact, on-chip integration. Thus, there is a need for an active nanophotonic device capable of electrically modulating emissivity with reconfigurable spatial patterns.

Here, we present an electrically addressable plasmonic metasurface for the dynamic modulation of coherent and incoherent mid-IR radiation across a 2D array. First, we demonstrate the principle of our design by achieving tunable diffraction of coherent mid-IR light in one dimension (Fig. 1a). Then, we introduce a straightforward modification of the electrical interconnects and optical antennas that allows tunable diffraction of mid-IR light in two dimensions and for any polarization of light (Fig. 1b). Finally, by heating our device, we demonstrate spatially resolved dynamic emissivity modulation of thermally emitted light across our metasurface. This work presents a framework for designing active metasurfaces for the steering of coherent light and modulation of incoherent, thermally emitted light in the mid-IR.

**Results**

**Design principles for an active plasmonic metasurface in the mid-IR**

Transparent conducting oxides (TCOs) are degenerately doped wide bandgap semiconductors which are typically transparent in the visible and near-infrared (near-IR). Some examples of common TCOs are indium-tin-oxide (ITO), aluminum-doped zinc oxide (AZO), gallium-doped zinc oxide (GZO), and cadmium oxide (CdO). The high density of charge carriers in TCOs is typically modeled as a "free electron gas" which determines the optical response of the material. The dielectric function of a TCO can be described by the Drude-Lorentz model which takes the form:

$$\varepsilon(\omega) = \varepsilon_\infty - \frac{\omega_p^2}{\omega^2 + iw\Gamma} \qquad (E1)$$

where $\varepsilon$ is the frequency-dependent complex permittivity, $\varepsilon_\infty$ is the high-frequency permittivity, and $\omega$ is the angular frequency of incoming light. The plasma frequency, $\omega_p$, and damping coefficient, $\Gamma$, are given by:

$$\omega_p^2 = \frac{Nq^2}{\varepsilon_0 m_e^*} \qquad (E2)$$

and

$$\Gamma = \frac{q}{m_e^* \mu} \qquad (E3)$$

where $N$ is the bulk carrier concentration, $q$ is the elementary charge, $\varepsilon_0$ is the permittivity of free space, $m_e^*$ is the effective mass of an electron, and $\mu$ is the electron mobility. An important feature of the Drude-Lorentz model is the existence of a frequency where the real part of the permittivity crosses zero, named the epsilon-near-zero (ENZ) point. At this frequency, the optical field becomes delocalized in the film, causing strong absorption and many other interesting optical phenomena (*26–29*). The ENZ frequency is determined by four material parameters in the Drude-Lorentz model: $N$, $\varepsilon_\infty$, $\mu$, and $m_e^*$. Among these, the carrier concentration, $N$, is the easiest to control experimentally and is often used to shift the ENZ frequency of a TCO film. To date, ITO has been the most studied TCO material and is used as the active material in this work.

ITO (typically Sn-doped $In_2O_3$) can be deposited via many different techniques such as sputtering, evaporation, atomic layer deposition (ALD), and sol-gel processing (*30*). Of these deposition methods, radiofrequency (RF) magnetron sputtering is the most common due to a favorable tradeoff between deposition rate, film quality and uniformity, film tunability, and capability for large area processing. A sputtering target containing a 90/10 weight % mix of $Sn_2O_3$ and $In_2O_3$ is typically used and sputtered in an environment of Ar and $O_2$ (*31*). During deposition, the $O_2$ flow rate and substrate temperature can be used to control the carrier concentration of the film (*32*). Importantly, post-deposition annealing in various oxygen environments can also provide tunability, which we leverage here (*33*).

The free carriers in ITO are contributed from indium donors and oxygen vacancies. While the concentration of indium donors is intrinsic to the film, the density of oxygen vacancies can be changed after deposition. In most cases, the carrier concentration of ITO increases as the density of oxygen vacancies increases. Thus, ITO films deposited with a low (high) oxygen flow rate tend to result in films with a high (low) oxygen vacancy concentration and thus a high (low) carrier concentration. This trend, however, only holds within a certain region of the complex deposition parameter space and is dependent on many other sputtering conditions such as substrate temperature, pressure, and power. Thus, it can be difficult to obtain a wide range of carrier concentrations by only controlling oxygen flow rate during deposition. Typically, achievable carrier concentrations are in the range of $1–5x10^{20}$ $cm^{-3}$, corresponds to ENZ frequencies in the near-IR. Post-deposition annealing in various oxygen environments can provide ENZ tuning over a wide spectral band (*33*), and Fig. 2a shows a visual schematic of how the concentration of oxygen vacancies in ITO can be manipulated by annealing in an oxygen-rich (e.g. air) versus an oxygen-poor (e.g. vacuum) environment. The left-most film in Fig. 2a shows an as-deposited ITO film which consists of a fixed concentration of oxygen vacancies ($V_O^{\cdot\cdot}$) and oxygen interstitials ($O_i''$). During an anneal in air, there is a net movement of oxygen into the film which fills and eliminates oxygen vacancies, resulting in a decrease of the overall carrier concentration (upper right panel of Fig. 2a). In contrast, annealing in vacuum will cause oxygen interstitials to exit the film, leaving behind oxygen vacancies, thus increasing the carrier concentration (bottom right panel of Fig. 2a).

As the ENZ frequency of ITO is typically in the near-IR, previous studies of ITO-based plasmonic metasurfaces for dynamic beam-steering operate in this wavelength band (*34–37*). The device design in these demonstrations could be applied to other wavelength ranges such as the mid-IR and longwave-IR (LWIR), but a significant ENZ wavelength red-shift of the ITO is required. Other materials such as $n^+$-doped InAs have been explored for mid-IR devices (*21*), but the requirement for high quality growth (often via molecular beam epitaxy) (*38*) and precise doping of InAs can make it difficult to integrate into complex optoelectronic devices. In this work, we develop a process that significantly decreases the carrier concentration of ITO films following deposition. This approach allows us to precisely tune the ENZ wavelength and provides flexibility in the design of other layers in our device. Figure 2b plots the real and imaginary permittivity components (extracted by fitting to ellipsometric measurements) of an as-deposited ITO film in our RF magnetron sputterer (left plot) and after a 4.5 hour anneal in air (right plot). These two measurements highlight our capability to shift the ENZ wavelength from approximately 4 $\mu$m to 13 $\mu$m. By annealing at relatively low temperature (150 °C), the diffusion of oxygen in and out of the film is very controlled such that it takes many hours or days to reach equilibrium. The final carrier concentration is highly dependent on the anneal time and allows us to precisely control the final properties of our ITO films. Thus, this constitutes controllable method to achieve a wide range of ENZ wavelengths in ITO films, widening the range of possible frequency bands for the operation of active ITO-based optical devices.

**Active resonance modulation**

Figure 3a presents a cross-section of the active metasurface device. The uppermost Au layer consists of antennas with a period ($\Gamma_{ant}$), width ($w_{ant}$), and height ($h_{ant}$) of 1400 nm, 925 nm and 40 nm, respectively. These antennas sit on top of uniform layers of ITO ($h_{ITO}$ = 50 nm) and $Al_2O_3$ ($h_{Al_2O_3,2}$ = 160 nm). A layer of gold electrodes ($h_{Au,1}$ = 80 nm) surrounded by $Al_2O_3$ enable electrical tuning of a *p*-polarized gap plasmon resonance hosted between the antennas and electrodes. Each electrode spans three antenna periods, with voltages applied between the bottom Au electrodes and top ITO layer to modify the carrier concentration of the ITO. This results in a local modulation of the reflection coefficient above each electrode, as schematically depicted in Fig. 3a. Importantly, the pitch of the addressable electrodes ($\Gamma_{electrode}$) remains subwavelength to suppress unwanted diffraction orders and allow for the steering of light over a wide field of view. The electrodes are separated from a back gold film ($h_{Au,2}$ = 80 nm) with a layer of uniform $Al_2O_3$ ($h_{Al_2O_3,1}$ = 200 nm) to prevent radiation from being transmitted through the device. The metasurface is wirebonded to an interposer printed circuit board (PCB) that interfaces with a set of custom-built programmable PCBs capable of applying voltages between -30 V and 30 V to 32 individual channels, each corresponding to a gold electrode. The electrodes corresponding to the 32 independent voltage channels are repeated three times across the metasurface aperture to form the full device, which measures 400x400 $\mu m^2$. In Fig. 3b, a scanning electron micrograph shows the antennas patterned above the addressable electrodes and the inset provides a magnified view of a single electrode with three corresponding antennas. Supplementary Materials (Figs. S1–S3) provide additional optical and scanning electron microscope (SEM) images of our active metasurfaces.

We begin by characterizing the electrostatic tuning of the resonance of the metasurface. When a negative voltage is applied to the addressable gold elements, the ITO becomes depleted of charge carriers near the interface ITO/$Al_2O_3$. This in turn modulates the gap plasmon resonance hosted between the gold antennas and the addressable gold electrodes below, enabling local control of scattered light. We simulate the reflectance of this structure with a finite-difference time domain (FDTD) solver (Lumerical, Inc.) under *p*-polarized illumination at normal incidence for wavelengths between $\lambda = 4-5$ $\mu m$. Figure 3c shows the simulated reflectance spectra for a series of assumed carrier concentrations in a 10 nm region of ITO to approximate a depletion layer. As discussed above, the optical properties of ITO are captured by the Drude-Lorentz model, and based on electrostatic simulations, we assume that the bottom 10 nm of the ITO layer (adjacent to the bottom $Al_2O_3$ layer) is modulated when in depletion. Justification of this thickness and further discussion is provided in the Supplementary Materials (Fig. S4). Indeed, because the field of the gap plasmon is highly concentrated in this region, the simulated spectrum shifts as the carrier concentration is reduced. This is in contrast to prior work with ITO-based active metasurfaces where the carrier concentrations are typically much larger and only a ~ 1 nm layer is modulated with carrier *accumulation*. Since our carrier concentration is an order of magnitude lower, the volume of the *depletion* layer is roughly an order of magnitude greater and results in strong modulation.

The measured reflectance spectra also shift appreciably with applied negative voltage (Fig. 3d). A detailed description of the measurement setup is given in the Supplementary Materials (Fig. S5). At 0 V, the resonance is centered near $\lambda = 4.95$ $\mu m$. We are limited to measuring up to $\lambda = 5$ $\mu m$ due to the tuning range of our laser (MIRcat, Daylight Solutions Inc.). The resonant frequency of the fabricated device likely differs from that of the simulated one because of differences in the carrier concentration of the ITO from what is assumed in simulation and measured prior to depositing the gold antennas.

Despite this, on resonance the reflectance increases from approximately 2% to 9% when applying -51 V to the device.

**Active steering in one dimension**

In the previous section, we studied the ability of our metasurface to modulate incident light uniformly across its aperture. Now, we use the individually addressable electrodes to apply spatially varying modulation patterns that deflect incident light to different directions. The addressing scheme is demonstrated in Fig. 4a, where distinct voltages can be applied to each of the 32 elements such that the reflectance associated with each group of three antennas can be modulated across the metasurface aperture. To demonstrate this functionality, we apply a series of binary grating patterns with different periods to symmetrically diffract light to three distinct angles. We define $P_0 = 3\times1.4$ $\mu$m as the period of the addressable electrodes (Fig. 3a), which sets the maximum diffraction angle achievable by our metasurface. Grating patterns with three periods ($P = 4P_0$, $P = 8P_0$, and $P = 16P_0$,) are applied to the metasurface with voltages of -60V and 0V on adjacent groups of electrodes (Fig. 3a).

To characterize the diffraction from the metasurface, we image the Fourier plane using a ZnSe ashperic lens for each grating pattern in Fig. 4a. The experimental setup is detailed in Supplementary Materials (Fig. S5). Figure 4b shows the measured Fourier plane images where we observe clear, symmetric diffraction orders between 4° and 18° for a wavelength of $\lambda = 4.95$ $\mu$m. We have subtracted a background frame without any illumination. Additionally, the images are normalized to the peak $0^{th}$ order intensity. Offset line cuts of these images along $\theta_y = 0°$ are shown in Fig. 4c. The intensity of the steered light is approximately 5% of the $0^{th}$ order because the resonance modulation is amplitude dominated with minimal phase shift between adjacent elements. Due to the limited phase shift, we do not expect to strongly suppress the $0^{th}$ order and we estimate the total diffraction efficiency to be near 0.3%. The primary challenge in achieving strong phase modulation in our devices is the sensitivity of the ITO film to its environment due to its exceedingly low carrier concentration (and thus higher oxygen content). In a device with strong phase shift, the $0^{th}$ order can be suppressed and light can be steered asymmetrically. However, the overall efficiency of the steered light may be comparable to the case of pure amplitude modulation because phase modulation in plasmonic metasurfaces occurs near critical coupling, necessarily introducing stronger absorption (*35*).

We next study the temporal response of the metasurface by driving all electrodes with square wave signals with frequencies between 1 and 400 kHz. We measure the reflected $0^{th}$ order signal from the metasurface on a HgCdTe photodetector (InfraRed Associates Inc., MCT-5-N-0.25) coupled to a preamplifier (Infrared Systems Development, MCT-1000). In Fig. 4d, the left panel shows the measured reflected signal when driving the metasurface with square waves of $f = 10$ kHz, 50 kHz, and 150 kHz. In the right panel, the relative modulation depth is plotted as a function of the driving frequency, and we find that the 3 dB bandwidth is approximately 155 kHz. Although free-space ITO-based electro-optic devices have been demonstrated to be capable of modulation frequencies in the GHz range (*39*), the preamplifier used in our experiments has a bandwidth of 150 kHz, which ultimately limits the measurable frequency range. Thus, the metasurface itself is not limiting the measured modulation bandwidth and can likely be modulated at much higher frequencies than reported here.

**Polarization-independent 2D control of diffraction and thermal emission**

A fundamental challenge with 2D modulation of active metasurfaces is the increased complexity of the driving electronics and electrical layout to address individual pixels (*40, 41*). Recently, it has been proposed that a perimeter-addressed array of crossed electrodes (similar to passive matrix driving schemes in displays) can be used to achieve 2D beam steering in an active metasurface consisting of scattering elements with a strong phase response (*42*). Inspired by this work, we introduce a simple modification to our metasurface architecture by patterning ITO layer to create addressable electrodes running orthogonal to the gold electrodes below. A schematic top-view of this configuration is shown in Fig. 5a. The resonant antennas are now circular (producing a polarization-independent resonance) and sit in 12x12 arrays atop the regions where the ITO and gold strips overlap. The antennas still maintain a period of 1.4 $\mu$m, but between each group of 12x12, there is a 400 nm gap introduced to better isolate the electrodes electrically. Thus, the period of the gold and ITO electrodes is 17.2 $\mu$m. These antenna groups are modulated by applying voltages to the corresponding row and column electrodes. Now that there are additional addressable electrodes along the y-direction, the 32 programmable channels are split into two groups of 16, routed to separately control the gold and ITO electrodes that span the entire 275 x 275 $\mu m^2$ device aperture.

First, we study the resonance modulation of this device by measuring its reflectance spectrum with a uniform voltage applied across the aperture. In Fig. 5b, the experimentally measured reflectance spectrum is shown at normal incidence for orthogonal incident polarizations with 0V and -60V applied across the top $Al_2O_3$ to deplete the ITO. We find that the resonance is centered near $\lambda$ = 4.4 $\mu$m and observe strong modulation of the reflection spectrum for both polarizations. Small differences in the reflectance for orthogonal polarizations are attributed to the patterned ITO and gold electrodes slightly breaking the symmetry that would be present in the case of a uniform antenna array. Next, we probe the ability of our device to steer light in orthogonal directions. With an identical Fourier plane imaging setup as in the previous section, we capture images at $\lambda$ = 4.4 $\mu$m with no voltage applied, a grating pattern applied to the gold electrodes (*x*-axis), and a grating pattern applied to the ITO electrodes (*y*-axis). The resulting images for *y*-polarized light are shown in Figs. 5c-e and confirm that our device can steer reflected light in orthogonal directions, albeit with low efficiency. The $0^{th}$ order is saturated on the camera to see more clearly the first order diffraction spot in each voltage configurations. The corresponding images for *x*-polarized light are included in the Supplementary Materials (Fig. S6) and show near-identical steering. Based on the reflectance spectrum of Fig. 5b exhibiting less relative amplitude change compared to the 1D device, it is expected that the diffraction efficiency should be lower. However, as will be discussed later, the higher overall reflectance translates to a greater relative modulation of emissivity. Another reason for low diffraction efficiency is that some adjacent electrodes were found to be electrically shorted due to fabrication imperfections and needed to be disconnected. As a result, the grating pattern is not perfectly periodic, which has a significant effect on the diffraction efficiency and spot size.

The fact that our metasurface operates in the mid-IR affords us the opportunity to explore its potential to spatially modulate thermally emitted radiation. Kirchoff's law of thermal radiation states that the emissivity of an object is equal to its absorptivity, and there have been recent demonstrations of metasurfaces that modulate emissivity by tuning their resonances (*20*, *21*, *43*). Additionally, previous works have used optical pumping to locally modulate emissivity (*25*). However, to the best of our knowledge, there has been no demonstration of a device capable of active spatial patterning of thermal emissivity using electrical modulation. To investigate this in our metasurface, we image the metasurface (setup shown in the Supplementary Materials Fig. S5) and apply the same three grating patterns as in Figs. 5c-e in the absence of laser illumination. The metasurface reaches a temperature of approximately 50 °C due to the heat of the driving electronics beneath it. Without an additional heat source, in Fig. 5f we clearly observe the metasurface aperture emitting radiation with an intensity greater than that of the surrounding materials when no voltage pattern is applied. Then, by applying horizontal and vertical grating patterns to the ITO and gold electrodes, visible sets of orthogonal lines in each direction are observed (Figs. 5 g,h). We note that the image in Fig. 5h was taken from a different metasurface than the images in Figs. 5f,g due to inadvertent electrical damage to the latter. The absolute emissivity is obtained by referencing the measured camera counts to a soot sample, which we assume approximates a perfect blackbody emitter. These lines are a direct result of an emissivity change due to the resonance modulation changing the absorptivity of the metasurface at constant temperature and are not caused by local heating. Missing lines are the result of the shorted electrodes being disconnected, as mentioned above. Thermal emission is spectrally broadband so we must consider the origin of the thermal signatures captured. Our camera is only sensitive to wavelengths from 1-5 $\mu$m, and the Planck blackbody spectrum of a 50 °C distribution is centered near $\lambda = 16$ $\mu$m such that any absorptive resonance at shorter wavelengths contribute minimally to the emissive signal. In the Supplementary Materials (Fig. S7), we present the simulated reflectance spectrum spanning our camera's sensitivity window for two different ITO carrier concentrations and confirms that the primary resonance responsible for emission is the magnetic dipole gap plasmon considered in our devices. Thus, we conclude that the source of the patterned thermal emission is a change in emissivity due to modulation of the optical resonance demonstrated in Fig. 5b.

**Discussion**

In this work, we have introduced a novel active metasurface platform for controlling coherent and incoherent mid-IR radiation in one and two spatial dimensions. We rely on post-fabrication thermal processing to tune the bulk ITO carrier concentration and precisely control its complex dielectric function to achieve gap plasmon resonances between wavelengths of 4–5 $\mu$m. In a 1D grating device, we have demonstrated symmetric beam steering with addressable elements of subwavelength spacing based on amplitude modulation across the metasurface aperture with diffraction efficiencies comparable to existing approaches. We then extended this platform to show the diffraction of light in two orthogonal directions with a perimeter addressing scheme that scales well with aperture size. Finally, by heating our device, we demonstrated electrically patterned thermal emissivity across the metasurface aperture at constant temperature.

The highest diffraction efficiency of our device was limited to approximately 0.3% due to our use of amplitude modulation. While this is efficiency low, it should be noted that it is comparable to what may be achieved with in an ITO-based plasmonic device with large phase modulation as this device would inherently have lower reflectance. Compared to previous demonstrations in the near-IR (*35–37*), our 1D steering exhibits larger reflected power in the $0^{th}$ order, but similar reflected power in the diffracted orders of interest. Furthermore, higher reflectance implies less absorption and thus less heating, meaning it is better suited to high-power illumination.

In the future, further optimization of our metasurface fabrication can allow for stronger phase modulation. This would enable highly directive and asymmetric steering of an incident beam in two spatial dimensions via perimeter addressing, which is not possible with pure amplitude modulation. Additionally, our work may ultimately lend itself to coherent control of thermal emission, whereby a nonlocal mode can be engineered to extend the spatial coherence of the emitted thermal radiation and couple adjacent scattering elements. We suspect that our results may open new opportunities for active devices in the mid-IR, a spectral region often overlooked yet replete with exciting applications.

**Materials and Methods**

**Device fabrication**

Fabrication began on a two-inch <100> silicon wafer with a 1 $\mu$m thermally grown oxide. The back gold layer was patterned using electron-beam lithography (EBL), deposition using electron-beam (e-beam) evaporation of 5 nm Ti (for improved adhesion) and 80 nm Au, then liftoff in Remover PG. The EBL system used was the Raith electron beam pattern generator (EBPG) 5000+ and the e-beam evaporator was the Kurt J. Lesker Labline Electron Beam Evaporator. Next, we patterned 200 nm of $Al_2O_3$ via atomic layer deposition (ALD) using a shadow mask. The ALD system used was the Ultratech/Cambridge NanoTech Plasma Enhanced ALD System Fiji G2. After the $Al_2O_3$ deposition, the whole wafer was annealed on a hotplate at 200 °C in air. Next, another gold layer was patterned using EBL, e-beam evaporation of 5nm Ti and 80 nm Au, then liftoff in Remover PG. This layer formed the bottom electrical interconnects for both 1D and 2D devices. After this, another masked ALD deposition of 160 nm $Al_2O_3$ was deposited via ALD. Next, 50 nm of indium tin oxide (ITO) was patterned using EBL, sputtering using an AJA International RF magnetron sputterer with a deposition pressure of 3 mTorr and RF power of 100 W while flowing 20 sccm pure Ar and 20 sccm $Ar/O_2$ in a 90/10 weight % mixture, then liftoff in Remover PG. For the 1D case, the ITO was patterned into a continuous film across the active metasurface area. For the 2D case, the ITO was patterned into individual strips to allow for separate biasing. At this point, we then performed multiple anneals our sample at 150 °C in air and vacuum until we measured the desired carrier concentration of our ITO film. Annealing in air decreased carrier concentration and annealing in vacuum increased the carrier concentration. We characterized the carrier concentration through ellipsometric measurements and Drude-Lorentz model fits with a wavelength range from 1.5 $\mu$m - 40 $\mu$m. Once the desired ITO carrier concentration was obtained, we finally patterned Au antennas using EBL, e-beam evaporation, and liftoff in Remover PG. For this final step, all baking steps were performed at 100 °C to prevent altering the ITO properties. Additionally, we did not use a Ti adhesion layer to avoid any reaction between the Ti and ITO film.

**Measurements**

All optical measurements were performed using a home-built setup. Supplementary Materials (Fig. S5) describes this setup in more detail.

**Device Simulations**

All optical simulations were performed using Ansys Lumerical finite-difference time-domain (FDTD) Solutions. Charge carrier simulations were performed using the Ansys Lumerical CHARGE module (see Supplementary Materials Fig. S4).

**Acknowledgments**

**General**
We gratefully acknowledge the critical support and infrastructure provided for this work by The Kavli Nanoscience Institute at Caltech.

**Funding**

This work was supported by the Air Force Office of Scientific Research Meta-Imaging MURI Grant No. FA9550-21-1-0312 (JS, MK and HAA), Caltech Space Solar Power Project (JS, MK and HAA), and the Army Research Office Grant No. W911NF-22-1-0097 (PP and AF). JS and PP acknowledge the support of the Natural Sciences and Engineering Research Council of Canada (NSERC) through the Postgraduate Scholarship – Doctoral program.

**Author Contributions**
JS, PP, AF, and HAA conceived the ideas for this research project. JS, with assistance from PP, performed the device fabrication. PP, with assistance from JS, performed the optical characterization. MK designed and built the 32-channel reprogrammable driving electronics and assisted with measurements. All authors contributed to the writing of the manuscript.

**Competing Interests**
The authors declare that they have no competing interests.

**Data and materials availability**
All data are available in the main text or the supplementary materials

**Figures and Tables**

Fig. 1: Active modulation of mid-IR radiation in two dimensions
Fig. 2: Engineering the properties of indium-tin-oxide (ITO)
Fig. 3: 1D active metasurface platform
Fig. 4: Reconfigurable diffraction in 1D
Fig. 5: 2D diffraction and emissivity modulation

**Supplementary Materials**

Supplementary Figure S1: Optical images of active metasurface devices
Supplementary Figure S2: Scanning electron microscope (SEM) images of 1D structure
Supplementary Figure S3: Scanning electron microscope (SEM) images of 2D structure
Supplementary Figure S4: Electron density distribution in ITO films of different bulk carrier concentration
Supplementary Figure S5: Experimental setup for active metasurface characterization
Supplementary Figure S6: Measured Fourier plane images with *x*-polarized illumination
Supplementary Figure S7: Spectral radiance analysis of our active metasurface
Supplementary Video S1: S1_1Dmetasurface_tunableDiffraction_0.5sPerAngle.mp4
Supplementary Video S2: S2_2Dmetasurface_tunableEmissivity_0.5sPerPattern.mp4

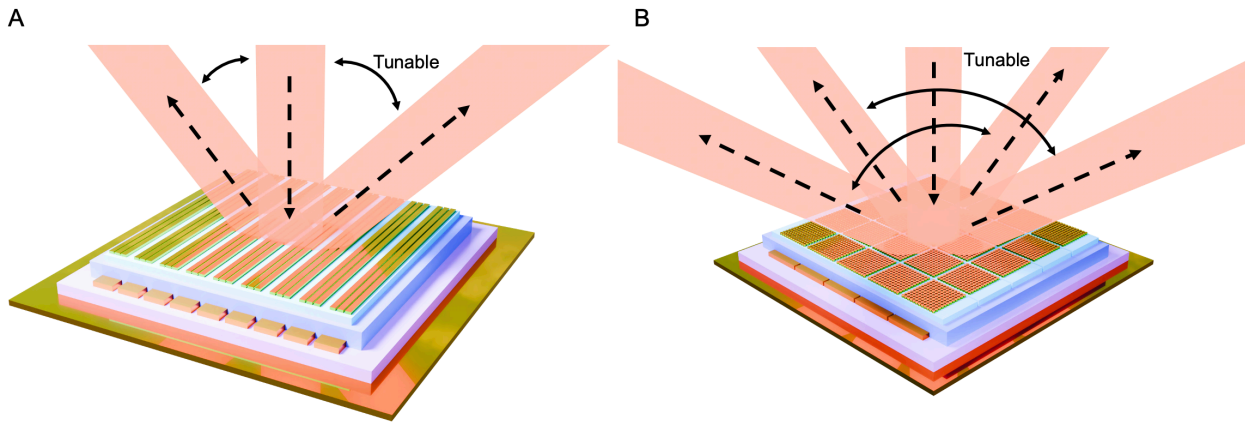

**Figure 1: Active modulation of mid-IR radiation in two dimensions. (A)** A 3D schematic of our active metasurface for the tunable diffraction in 1D of reflected coherent light in the mid-IR. **(B)** A 3D schematic of our active metasurface for the tunable diffraction in 2D of reflected coherent light in the mid-IR as well as the active modulation of thermally emitted light in 2D.

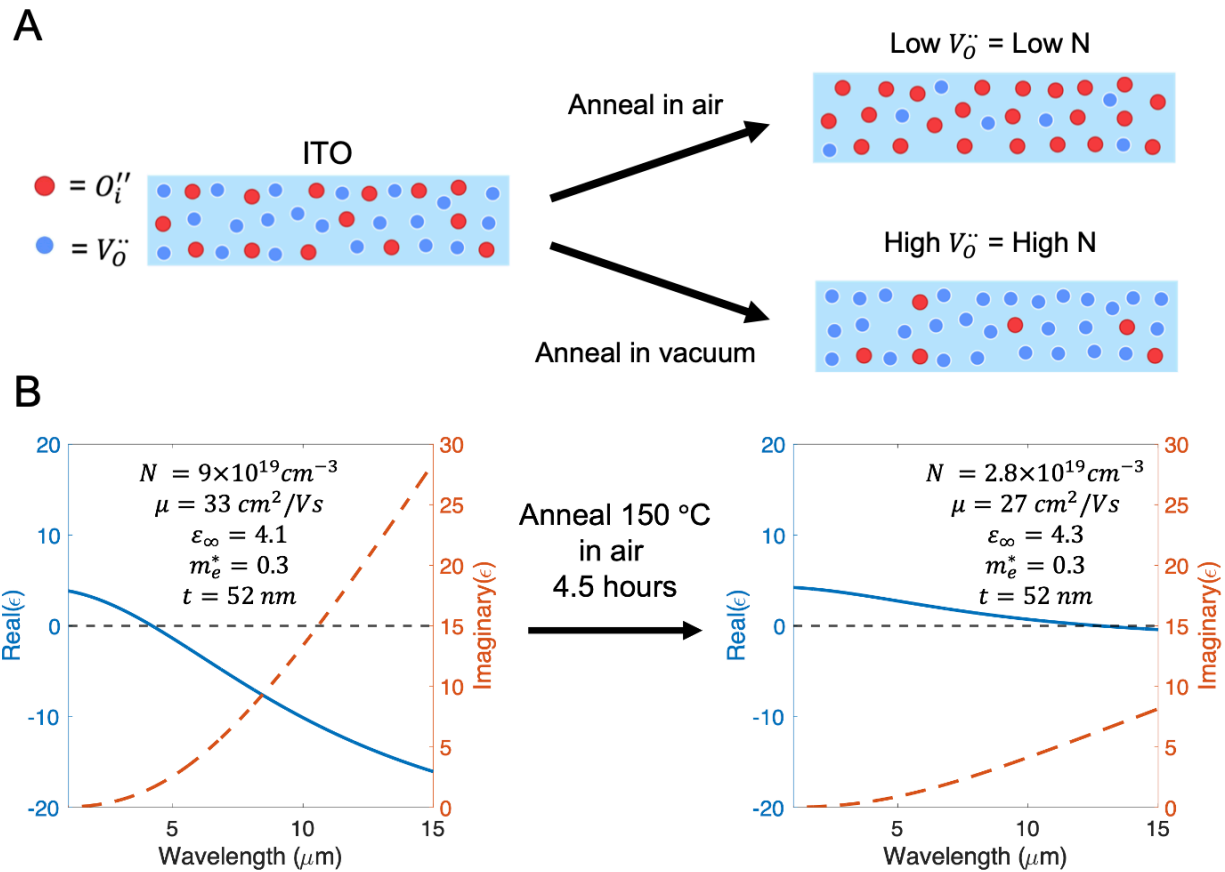

**Figure 2: Engineering the properties of indium-tin-oxide (ITO).** **(A)** ITO films can be modeled as a collection of oxygen interstitials, $O_i''$, (red circles) and oxygen vacancies, $V_O^{\cdot\cdot}$, (blue circles). The carrier concentration is proportional to the concentration of oxygen vacancies. After annealing in air (top right panel), the oxygen vacancy and carrier concentration is decreased. After annealing in vacuum (bottom right panel), the oxygen vacancy and carrier concentration is increased. **(B)** Experimentally measured permittivity, $\varepsilon$, of an ITO film directly after sputter deposition (left panel) and after a 4.5 hour 150 °C anneal in air (right panel). The epsilon-near-zero (ENZ) wavelength for each film is labelled and highlighted in blue.

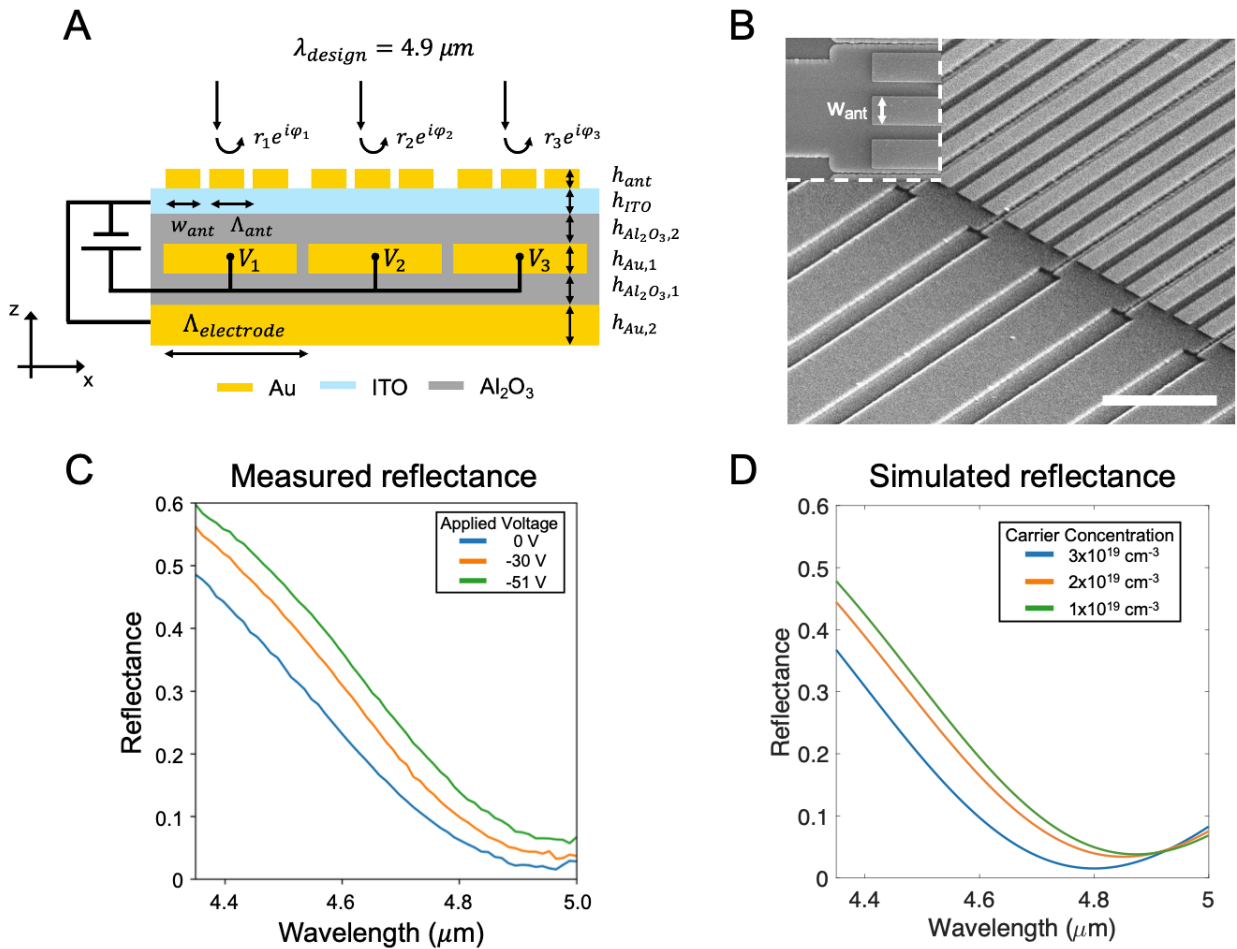

**Figure 3: 1D active metasurface platform.** **(A)** Cross section of the device showing the local modulation of the reflection coefficient with applied voltage. **(B)** Scanning electron micrographs of the fabricated metasurface. The inset shows a closer view of one set of three antennas. Scale bar for larger image: 5 $\mu$m. Labeled width in inset: $w_{ant}$ = 910 nm. **(C)** Simulated reflectance of the metasurface at normal incidence for *p*-polarized illumination for different assumed carrier concentrations. **(D)** Measured reflectance when applying three uniform voltages across the ITO and electrodes.

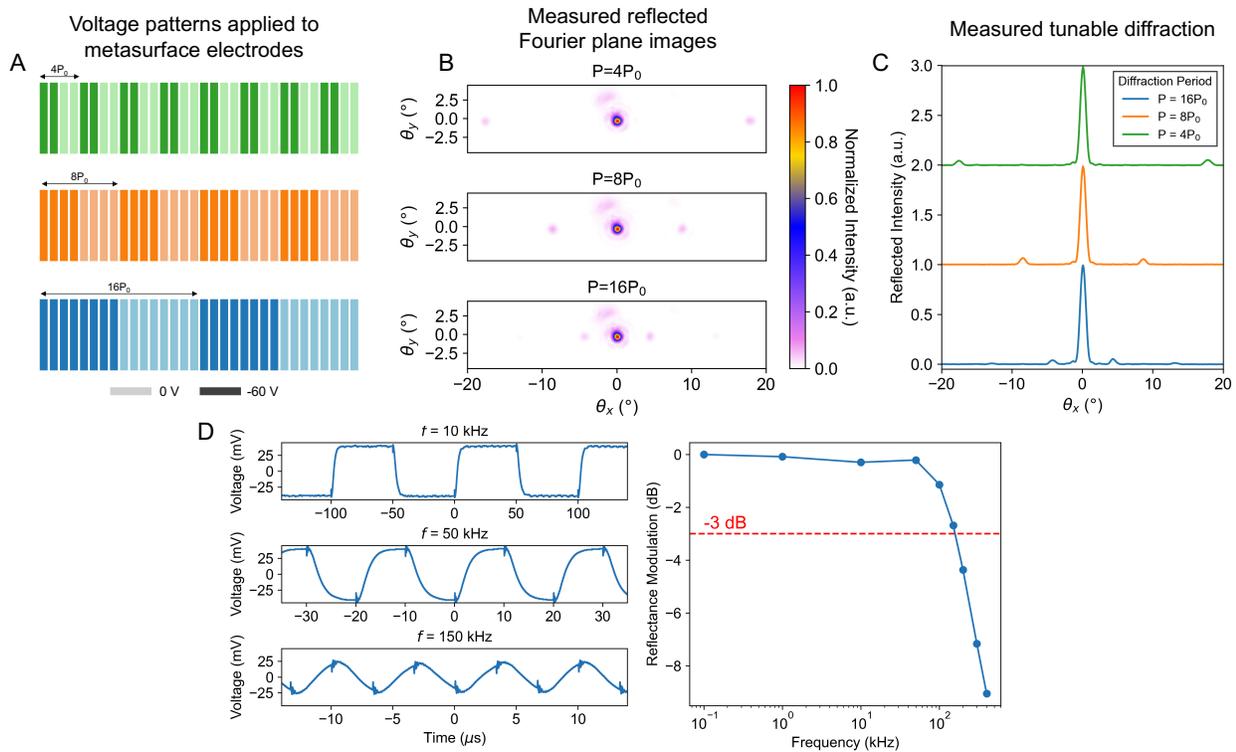

**Figure 4: Reconfigurable diffraction in 1D. (A)** Voltage patterns applied to the metasurface. The lighter and darker shades represent 0 V and -60 V, respectively. The period of the metasurface antennas is indicated as $P_0$. **(B)** Fourier plane images and **(C)** Line cuts through $\theta_y = 0°$ for the three voltage patterns in (A). The images are normalized to the $0^{th}$ order maximum in each of the three voltage configurations. **(D)** Measured $0^{th}$ order reflected signal of the metasurface when modulated uniformly with square waves of frequencies 10 kHz, 50 kHz, and 150kHz (left panel). Reflectance modulation depth as a function of driving frequency (right panel).

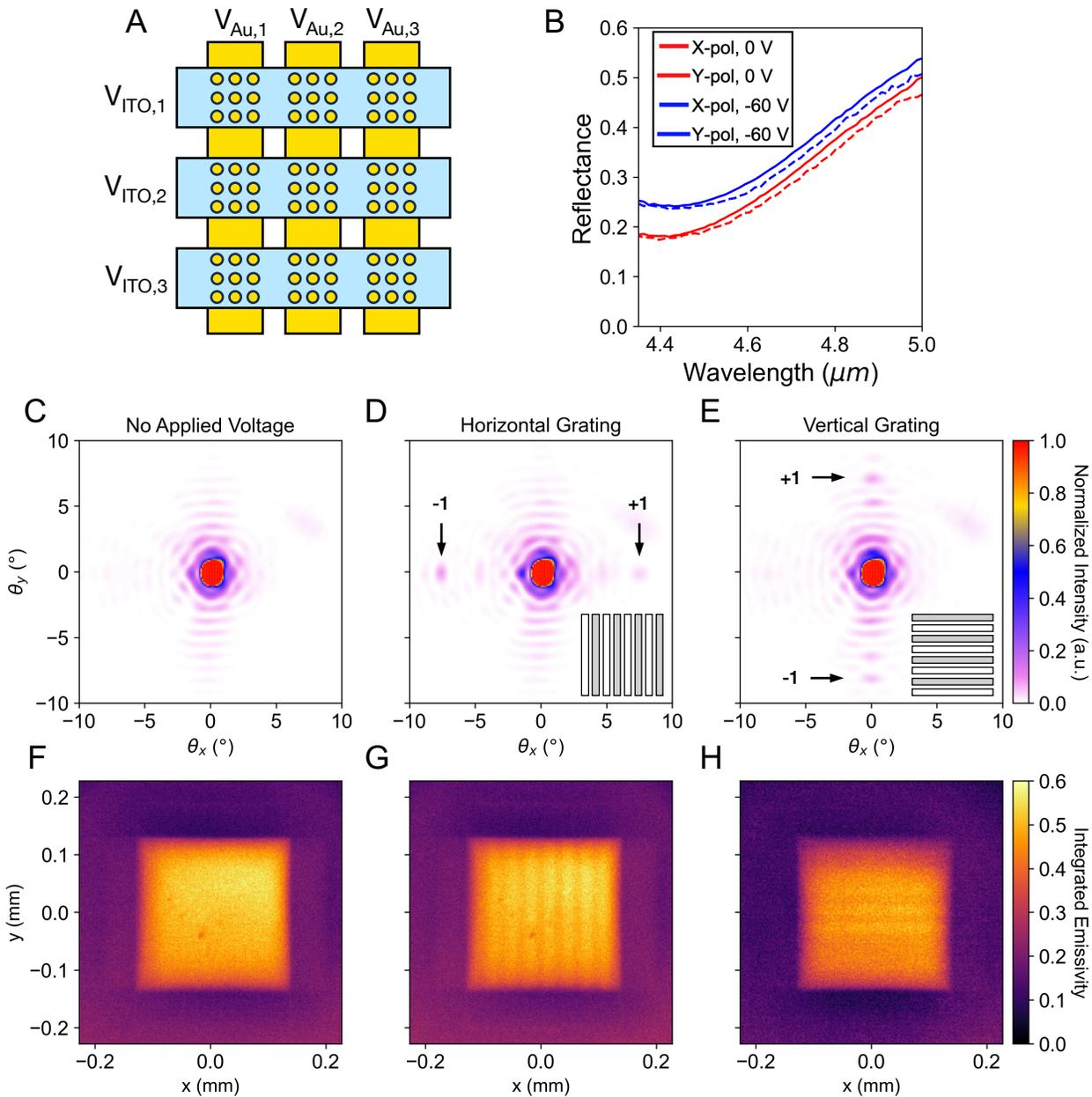

**Figure 5: 2D diffraction and emissivity modulation. (A)** Bird's eye view of the crossed ITO and Au electrodes enabling perimeter addressed modulation by applying voltages $V_{ITO,i}$ and $V_{Au,i}$. At the electrodes' intersections there are clusters of 12 x 12 circular Au antennas. **(B)** Measured reflectance spectrum for X- and Y-polarized illumination with 0 V and -60 V applied between the ITO and Au. The voltage pattern is uniform across the metasurface aperture. Measured Fourier plane images with Y-polarized illumination at $\lambda$ = 4.4 $\mu$m for **(C)** 0V on all electrodes, **(D)** alternating 0 V/-60 V on Au electrodes and 0 V for all ITO electrodes producing a grating running horizontally, and **(E)** 0 V/-60 V on alternating ITO electrodes and -60 V for all Au electrodes producing a grating running vertically. The inset diagrams in (D and E) show the orientation of the grating patterns. **(F-H)** Images of the emissivity across metasurface aperture captured in the absence of laser illumination for the same three voltage patterns as (C-E). The only source of heat is from the PCB upon which the metasurface is mounted, measured to be approximately 50 °C and the absolute emissivity is obtained by normalizing to a soot sample mounted identically to the metasurface.

# Supplementary Materials for

## Electro-optic modulation of coherent and incoherent mid-IR radiation in two-dimensional arrays

Jared Sisler†, Phillippe Pearson† *et al*.

*Corresponding author. Email: Harry A. Atwater (haa@caltech.edu)

**This PDF file includes:**
Supplementary Text
Figs. S1 to S7
References (1 to 3)

**Other Supplementary Materials for this manuscript include the following:**
Movies S1 to S2

**Supplementary Text**

Additional images of devices

We present additional images of our devices taken by optical microscope, optical camera, and scanning electron microscope. Figure S1 show optical microscope images of our 1D and 2D devices which show the active area in the center and the 32 electrical interconnects spreading out form the center. Figure S2 shows SEM images our 1D active metasurface device and Fig. S3 shows our 2D active metasurface device. In each case, one can see in more detail how the electrical interconnects address the grouped arrays of antennas which are precisely aligned on top of each contact. In both 1D and 2D devices, the spacing between the electrical contacts is very small, approximately 50 nm. This is most easily seen in Fig. S2f and Fig. S3e. This small electrical isolation allows us to not significantly perturb the periodicity of the antennas but makes our device sensitive to electrical shorting.

Electrostatic device simulations

We present the simulated electron charge density profile throughout our ITO films under an applied bias. Our simulations were completed in Ansys Lumerical CHARGE module and consist of a 50 nm film of ITO on top of a 200 nm film of $Al_2O_3$, sandwiched between to electrical contacts. The electrical permittivity of $Al_2O_3$ is assumed to be 9. The ITO electron mobility is assumed to be 30 $cm^2/(Vs)$, the effective mass 0.3, and the carrier concentration was changed between $3x10^{20}$ $cm^{-3}$ and $3x10^{19}$ $cm^{-3}$ to provide a comparison. These two values were selected as they represent a typical carrier concentration used in ITO-based active plasmonic devices in the near-IR ($3x10^{20}$ $cm^{-3}$) (*1*, *2*) and the carrier concentration used in this work ($3x10^{19}$ $cm^{-3}$). The voltage applied between the two electrical contacts was 50V (Fig. S4a) and -50 V (Fig. S4b) relative to the ITO/contact interface such that there was carrier accumulation/depletion in the ITO at the $Al_2O_3$/ITO interface. Figure S4a plots a 1D cross-section of the carrier concentration as a function of distance from the $Al_2O_3$/ITO interface for the two different bulk carrier concentrations of ITO under an applied bias of 50 V. The accumulation region widths are similar between the two carrier concentrations, with the $3\times10^{19}$ cm$^{-3}$ case being slightly larger its charge density is lower. For both bulk carrier concentrations, the accumulation region width is less than 5 nm. Figure S4b shows the charge profile when a voltage of -50 V is applied and the ITO is forced into depletion. Here, we can see the benefit of using charge depletion instead of accumulation for the lower carrier concentrations used in this work. Since the bulk charge is so much lower, the depletion region must be much larger in order to compensate for the voltage applied to the device. The depletion region associated with the $3\times10^{19}$ cm$^{-3}$ case is approximately 10 nm and is much larger than its associated accumulation width. Thus, by operating in depletion, we can make use of the electrical and optical modulation in a larger volume of ITO. For previous works in the near-IR, the difference between the accumulation and depletion widths are not as significant due to the larger bulk carrier density.

It should be noted that the simulations performed here assume the classical drift diffusion model and simply look for general trends between different bulk carrier concentrations. It has been reported that devices with confined dimensions (on the order of the electron wavelength) can be more accurately modeled using the Schrödinger-Poisson (SP) quantum model to calculate the carrier density (*3*).

Experimental optical setup for metasurface characterization

The experimental setup used to perform the measurements presented in this work is shown in Fig. S5. It is used in three configurations (1) for measuring reflectance spectra and high-speed modulation, (2) capturing Fourier plane images, and (3) imaging the metasurface plane for spatially-resolved emissivity measurements. A free-space tunable quantum-cascade laser (QCL, Daylight Solutions) is used to illuminate the metasurface for configurations (1) and (2). A lens L1 weakly focuses the beam to the back focal plane (BFP) of L2, which roughly re-collimates the beam as it illuminates the metasurface. The beam diameter is controlled by an adjustable iris. In configuration (1), L3 is removed from the beam path such that the light reflected from the metasurface is directly collected by a thermopile detector and the camera is conjugate with the metasurface plane. In configuration (2), L3 is in place which makes the camera conjugate with the BFP of L2. Then, to capture emissivity measurements are performed with the laser off and L3 removed from the optical path. This way, we directly capture the thermal emission in the metasurface plane.

Polarization dependence of diffraction in 2D

We present the measured diffracted light from our 2D active metasurface for *x*-polarized illumination. Figure S6 presents the Fourier plane images using all the same parameters as is presented in Fig. 5 of the main text, except for the incident illumination polarization being rotated 90°. The diffraction profiles look essentially identical to those in Fig. 5 of the main text for the opposite polarization and verify our device's insensitivity to polarization enabled by the circular antennas.

Analysis of measured radiance modulation

In Fig. 5 of the main text, we image an actively reprogrammable change in the emission of our metasurface across its aperture. Here, we show that this change in emission intensity corresponds to the change in emissivity at the designed resonance wavelength of approximately $\lambda$ 4.4 $\mu$m. In Fig. S7a, we plot the calculated spectral radiance of a blackbody at 50 °C, the measured temperature of our metasurface during measurement, using Planck's Law. The radiance peaks around $\lambda = 16$ $\mu$m and drops off sharply for wavelengths shorter than $\lambda = 10$ $\mu$m. The camera used in our experiments has a band edge corresponding to a wavelength of $\lambda = 5$ $\mu$m and we plot the corresponding responsivity window spanning $\lambda = 1$–$5$ $\mu$m as a shaded red box in Fig. S7a. Thus, due to the band edge of our detector, we can be sure that we are not detecting any change in emission from our device that may be occurring at wavelengths longer than $\lambda = 5$ $\mu$m. Additionally, as the radiance at 50 °C is very close to zero for wavelengths shorter than $\lambda = 1$ $\mu$m, we focus the rest of our analysis on the wavelength range from $\lambda = 1$-$5$ $\mu$m, where there is significant detectable radiation from our active device. Figure S7b plots the simulated reflectance from our device across this extended wavelength range for three carrier concentrations of the ITO depletion layer. (All assumptions in this simulation are identical to those used in the main text.) Here, we can see three distinct resonances. The designed resonance around $\lambda = 4.9$ $\mu$m corresponds to the magnetic dipole gap plasmon mode and is strongly modulated by the carrier concentration modulation. The resonance around $\lambda = 1.9$ $\mu$m corresponds to the electric dipole gap plasmon mode and is only slightly tuned by the carrier depletion. Finally, the narrow resonance around $\lambda = 1.4$ $\mu$m corresponds to the Rayleigh anomaly caused by the wavelength of light at normal incidence being equal to the antenna periodicity and is not significantly modulated. Next, we can calculate the expected radiance modulation from our device by first calculating the simulated emissivity. Because there is no transmission in our device, we can assume that the absorption is given by $1 - R$. Then, by Kirchoff's law of thermal radiation, we know that the absorptivity is equal to the emissivity. Thus, we plot the expected spectral radiance from our device in Fig. S7c by multiplying the spectral radiance in Fig. S7a by $(1 - R)$ from Fig. S7b at each carrier concentration. In this plot, we can more clearly see that the measured radiance is dominated by emission at wavelengths longer than $\lambda = 3$ $\mu$m. Additionally, there is slight modulation of the radiance at each simulated carrier concentration, which is strongest from wavelengths longer than $\lambda = 4$ $\mu$m. Finally, in Fig. S7d, we plot the absolute change in radiance between the maximum and minimum carrier concentrations from Fig. S7c. In this plot, we see that the largest change in measured radiance occurs at $\lambda = 4.5$ $\mu$m and significantly drops off for all other wavelengths. In conclusion, the active radiance modulation which we observe in the main text is a result of the modulation of the designed magnetic dipole gap plasmon mode, as expected.

Supplementary Videos

Three supplementary videos have also been included. Supplementary Video S1 ("1Dmetasurface_tunableDiffraction_0.5sPerAngle.mp4") shows the repeated tunable diffraction to the three different angles (shown in Fig. 4b of the main text), holding for 0.5 s at each angle. Supplementary Video S2 ("S2_2Dmetasurface_tunableEmissivity_0.5sPerPattern.mp4") shows the reconfigurable emissivity modulation between the patterns shown in Figs. 5f and g of the main text.

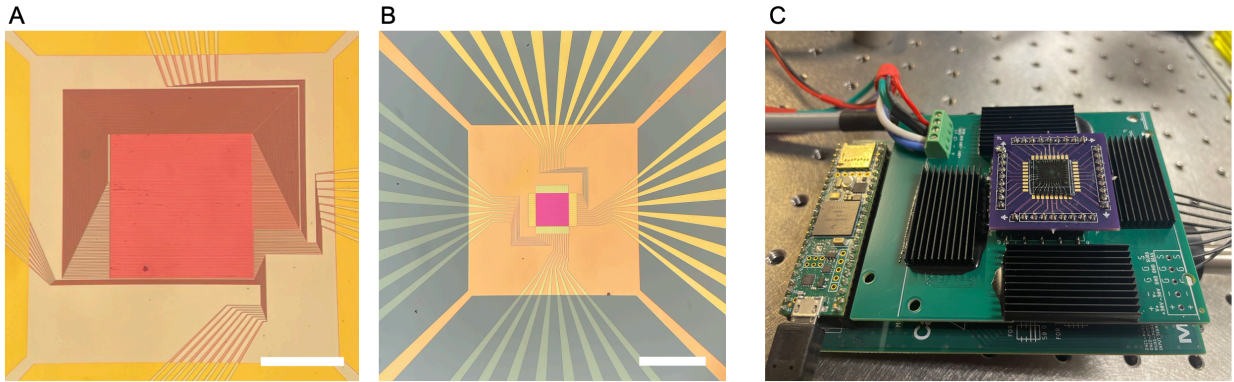

**Figure S1: Optical images of active metasurface devices. (A)** Optical microscope image of 1D device. Scale bar: 250 *µ*m. **(B)** Optical microscope image of 2D controllable device. Scale bar: 550 *µ*m. **(C)** Optical camera image of metasurface device connected to 32-channel driving PCB.

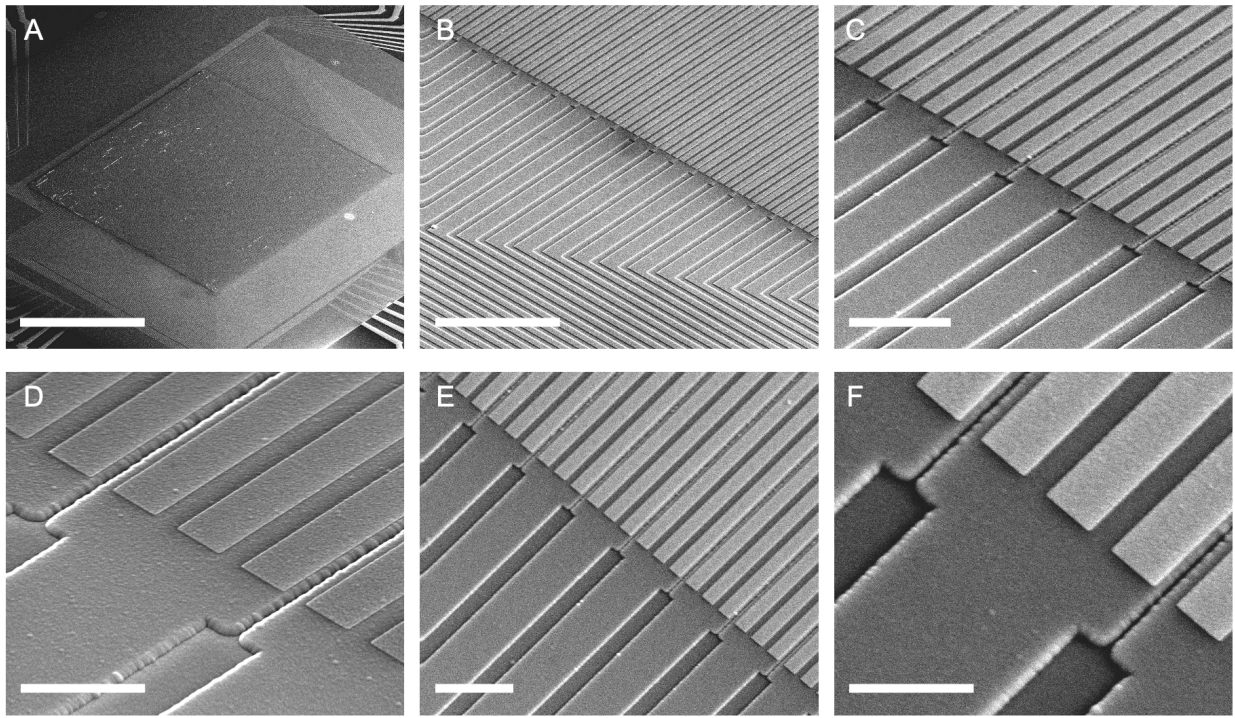

**Figure S2: Scanning electron microscope (SEM) images of 1D structure. (A-D)** Images taken at a tilt of 52°. **(E and F)** Images taken at normal incidence (no axis tilt). Scale bars for (A-F) are: 200 *µ*m, 20 *µ*m, 5 *µ*m, 2 *µ*m, 5 *µ*m, and 2 *µ*m, respectively.

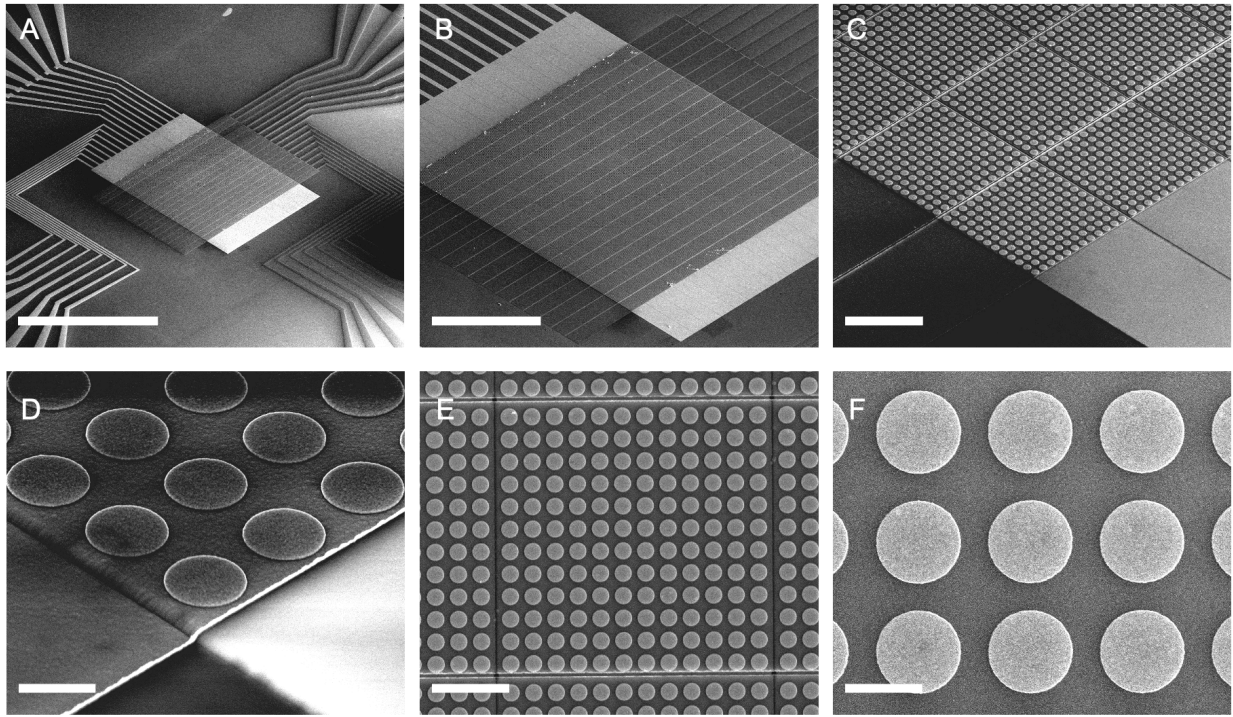

**Figure S3: Scanning electron microscope (SEM) images of 2D structure. (A-D)** Images taken at a tilt of 52°. **(E and F)** Images taken at normal incidence (no axis tilt). Scale bars for (A-F) are: 300 µm, 100 µm, 10 µm, 1 µm, 5 µm, and 1 µm, respectively.

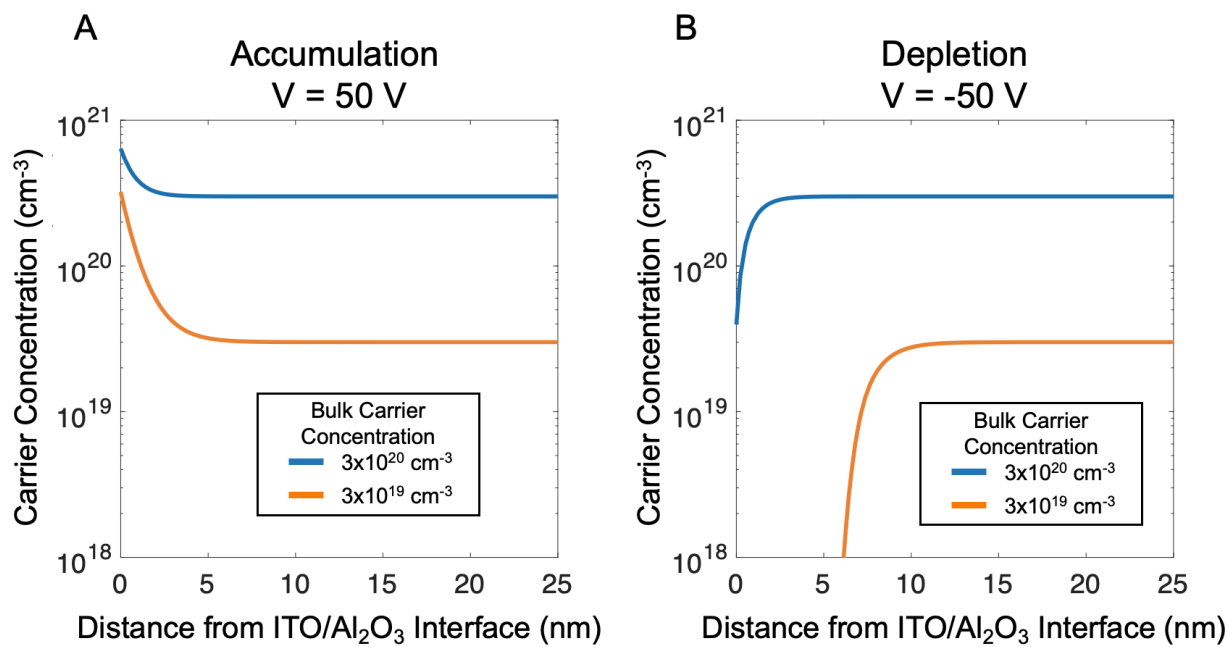

**Figure S4: Electron density distribution in ITO films of different bulk carrier concentration.** The simulated carrier concentration profile within 50 nm films of ITO with bulk carrier concentrations of $3 \times 10^{20}$ cm$^{-3}$ (blue line) and $3 \times 10^{19}$ cm$^{-3}$ (orange line). The carrier density is plotted as a function of the distance from the ITO/Al$_2$O$_3$ interface, for the first 25 nm of the ITO film, when a bias of 50 V **(A)** and -50 V **(B)** is applied.

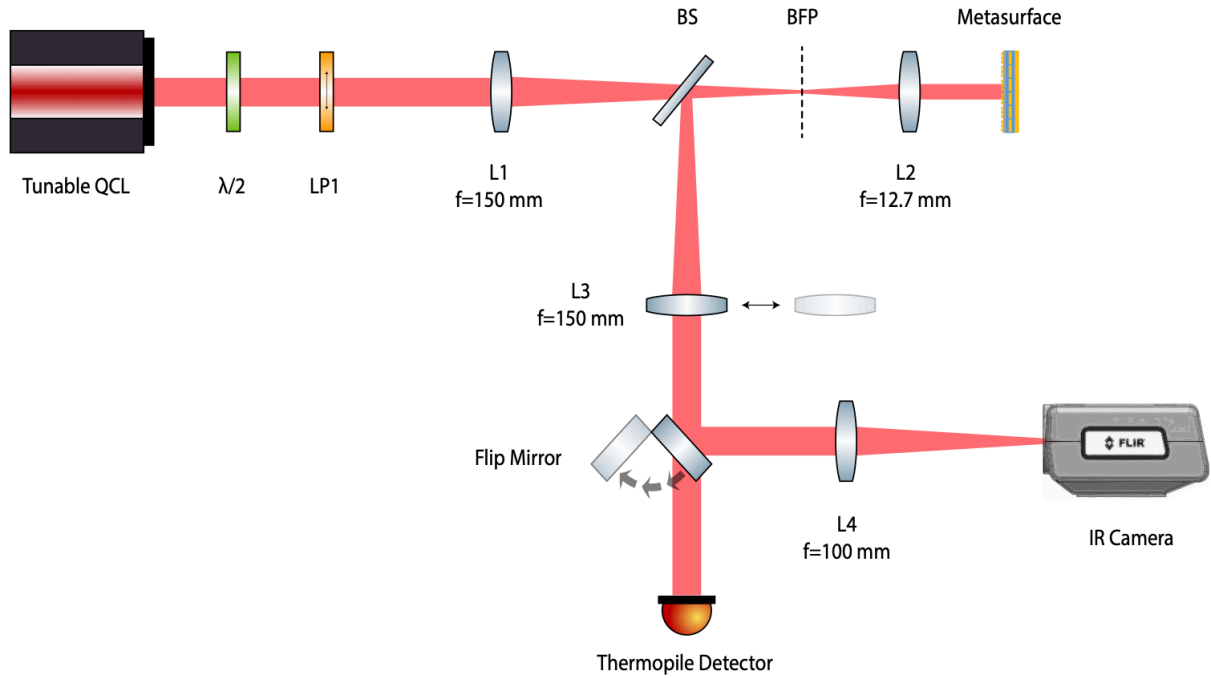

**Figure S5: Experimental setup for active metasurface characterization.**

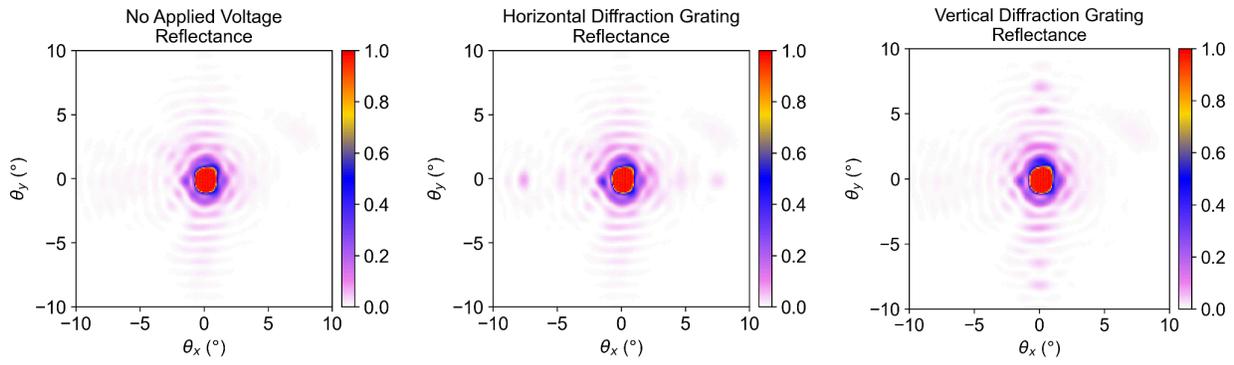

**Figure S6: Measured Fourier plane images with *x*-polarized illumination**. This is the *x*-polarized equivalent to the diffraction results presented in Fig. 5 of the main text for the 2D active metasurface.

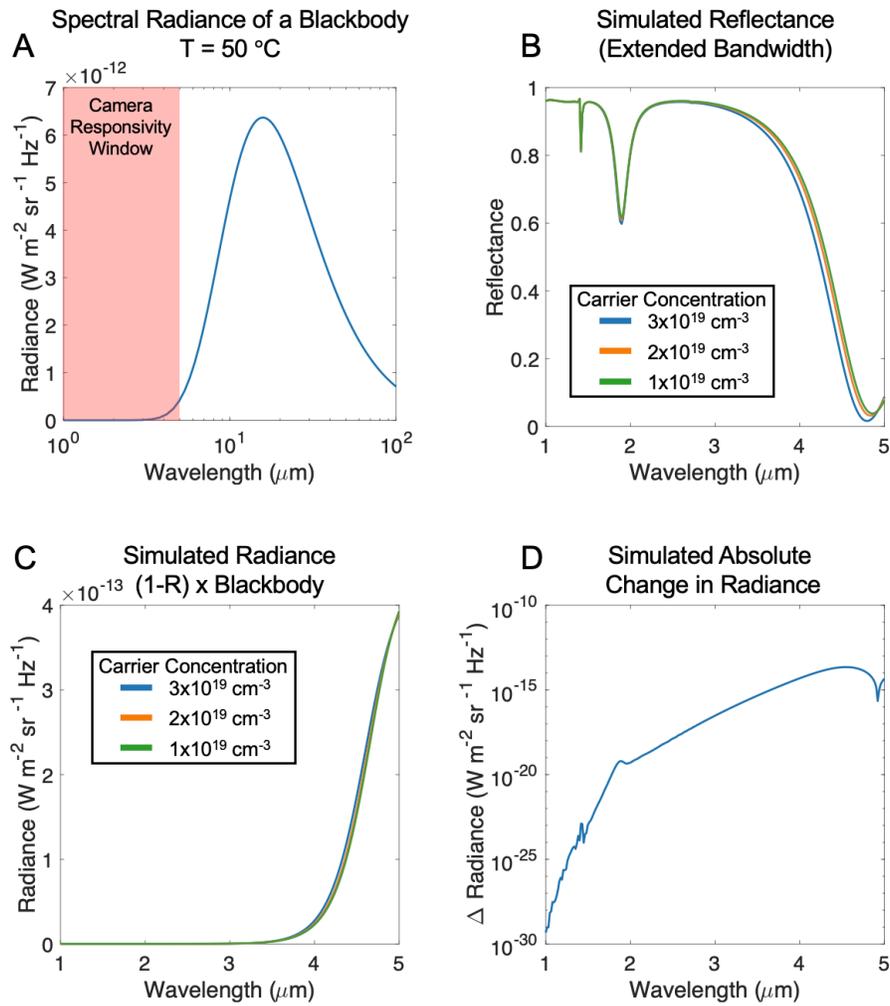

**Figure S7: Spectral radiance analysis of our active metasurface. (A)** Spectral radiance of a blackbody at T=50 °C calculated using Planck's Law. The responsivity window of the camera used in our measurements is shown in the red shaded region. **(B)** Simulated reflectance modulation of our active metasurface across a larger bandwidth for different assumed carrier concentrations in the ITO depletion layer. **(C)** Simulated total radiance modulation for our active metasurface. This plot was calculated by subtracting the simulated reflectance in (B) from 1 to obtain the absorption/emissivity, then multiplying by the radiance in (A). **(D)** Simulated absolute change in radiance in our active metasurface. This is obtained by subtracting the minimum radiance from the maximum radiance at each wavelength in (C).